\documentclass[aps, preprint, secnumarabic, floats, showpacs, nofootinbib, amssymb, amsfonts]{revtex4}
\usepackage[bookmarksnumbered, bookmarksopen, pdfstartview=FitH]{hyperref}

\begin{document}


\begin{titlepage}

\thispagestyle{empty}

\begin{center}

\hfill UMD-PP-04-043\\
\hfill \href{http://www.arxiv.org/abs/hep-th/0410157}{hep-th/0410157}\\

\vskip 2cm

{\LARGE \bf{${\cal M}$-Theory Vacua from Warped Compactifications on $Spin(7)$ Manifolds}}

\vskip 1cm

{\bf Drago\c{s} Constantin\footnote{\href{mailto:dragos@physics.umd.edu}{dragos@physics.umd.edu}}}

\vskip 0.5cm

{\em Department of Physics, University of Maryland\\
College Park, MD 20742-4111 USA}

\end{center}

\vskip 1cm

\begin{center} {\bf ABSTRACT } \end{center}

At the leading order, ${\cal M}$-theory admits minimal supersymmetric compactifications if the internal manifold has exceptional holonomy. Once we take into account higher order quantum correction terms in the low energy effective action, the supergravity vacua have to be deformed away from the exceptional holonomy if we want to preserve the supersymmetry of the solution. In this paper we discuss the $Spin(7)$ holonomy case. We derive a perturbative set of solutions which emerges from a warped compactification with non-vanishing flux for the ${\cal M}$-theory field strength and we identify the supersymmetric vacua out of this general set of solutions. For this purpose we have to evaluate the value of the quartic polynomial $J_0$ on a $Spin(7)$ holonomy manifold as well as its first variation with respect to the internal metric. We show that in general the Ricci flatness of the internal manifold is lost.

\vfill

\noindent{October 11, 2004}

\end{titlepage}


\setcounter{page}{2}


\section{Introduction}
\label{intro}

\setcounter{equation}{0}
\renewcommand{\theequation}{\thesection.\arabic{equation}}


Low dimensional compactifications of ${\cal M}$-theory have previously been discussed in the literature. In particular ${\cal M}$-theory compactifications on $Spin(7)$ holonomy manifolds\footnote{For a mathematical introduction into the subject of manifolds with exceptional holonomy the reader can consult the book by Dominic Joyce \cite{Joyce}.} lead to a minimal supersymmetric theory in three dimensions \cite{Shatashvili:1994zw, Papadopoulos:1995da}. In a previous work \cite{Becker:2003wb}, we have explicitly derived the low energy effective action that emerges from ${\cal M}$-theory compactification on $Spin(7)$ holonomy manifolds in the presence of non-zero background flux for the field strength. We have also determined the scalar potential generated for the moduli fields by the field strength flux and we have constructed an off-shell component formulation of the three-dimensional ${\cal N}=1$ supergravity coupled with an arbitrary number of scalar and $U(1)$ gauge fields. We concluded that most of the moduli fields are stabilized but the radial modulus remained unconstrained. The resulting three-dimensional theory is interesting because it can not be obtained by a reduction from a supersymmetric four-dimensional theory\footnote{The minimal supersymmetric theory in four dimensions compactified on $S^1$ produces a three-dimensional theory with ${\cal N}=2$ supersymmetry.}, thus one might understand the mechanism of ${\cal N}=1$ supersymmetry breaking in four dimensions by studying the three-dimensional theory with ${\cal N}=1$ supersymmetry. Also, because the string world-sheet is two-dimensional one expects to observe interesting phenomena upon compactification of string theory to two dimensions \cite{Gates:2000fj} and for this reason three-dimensional compactifications of M-theory are attractive. As well, the vanishing of the four-dimensional cosmological constant can be treated from a three-dimensional point of view as proposed by Witten in \cite{Witten:1994cg, Witten:1995rz} and exemplified in the three-dimensional case in \cite{Becker:1995sp}. Another strong reason to pursue a serious analysis of ${\cal M}$-theory on such a background is the close relation which exists between manifolds with $Spin(7)$ holonomy and manifolds with $G_2$ holonomy\footnote{${\cal M}$-theory compactified on manifolds with $G_2$ holonomy generates a minimal supersymmetric theory in four dimensions which is more appealing from a phenomenological point of view.}.

In general, due to membrane anomaly \cite{Vafa:1995fj, Duff:1995wd, Witten:1997md} and the global tadpole anomaly \cite{Sethi:1996es}, the compactification of ${\cal M}$-theory on eight-dimensional manifolds involves the presence of a non-vanishing flux for the field strength \cite{Becker:1996gj}. The supersymmetry imposes restrictions on the form of the field strength flux. In the $Spin(7)$ holonomy case the restrictions imposed to the flux were derived in \cite{Becker:2000jc}. It was later shown in \cite{Acharya:2002vs} and \cite{Becker:2002jj} that these constraints can be derived from certain equations which involve the superpotential\footnote{The external space is considered to be Minkowski.}
\begin{equation}
\label{w-conditions}
W=D_A W=0 \,,
\end{equation}
where $D_A W$ indicates the covariant derivative of $W$ with respect to the moduli fields which correspond to the metric deformations of the $Spin(7)$ holonomy manifold. We want to note that the compactness of the internal manifold was essential in the analysis performed in \cite{Acharya:2002vs} and \cite{Becker:2002jj}. In the present paper we restrict ourselves to manifolds with $Spin(7)$ holonomy which are smooth and compact\footnote{A compact manifold with $Spin(7)$ holonomy is simply connected. For details see \cite{Joyce}.}. However, as stated in \cite{Acharya:2002vs}, the result obtained using (\ref{w-conditions}) is valid for non-compact manifolds as well but the proof does not involve the above mentioned equations. The existence of a Ricci flat metric for such manifolds is not guaranteed as in the Calabi-Yau case therefore we will tacitly suppose that there are such metrics and we will perform all the derivations under this assumption. Even if we will be concerned only with compact manifolds which have Spin(7) holonomy we would like to mention a few papers such as \cite{Gukov:2001hf, Cvetic:2001pg} where non-compact examples of such manifolds have been constructed and analyzed. Also in \cite{Gukov:2002zg} aspects of topological transitions on non-compact manifolds with $Spin(7)$ holonomy and phase transitions have been considered. A more complete list of papers regarding singular manifolds with exceptional holonomy can be found in \cite{Acharya:2004qe} which is a recent review of the subject.

In the present paper we are interested in finding all the vacua generated by a warped compactification of ${\cal M}$-theory on compact eight-dimensional manifolds with $Spin(7)$ holonomy in the presence of a non-zero flux for the field strength. We will take into consideration all the known terms in the low energy effective action up to the order $\kappa_{11}^{-2/3}$, where $\kappa_{11}$ is the eleven-dimensional gravitational coupling constant. Not all the terms in the effective action are known to this order therefore we will need a criteria to consistently eliminate the contribution to the equations of motion that comes from these unknown terms. Terms like $F^2 R^3$ are known to appear in the $\kappa_{11}^{-2/3}$ order \cite{Strominger:1997eb} but they are suppressed in the large volume limit \cite{Peeters:2003ys}. Therefore the key ingredient would be to consider a perturbative series expansion in terms of a dimensionless parameter which is defined by the square of the ratio between the ``radius'' of the manifold\footnote{The ``radius'' of the manifold is nothing else but the characteristic length of the manifold.} and the eleven-dimensional Planck length. In a realistic scenario the average size of the internal manifold is much bigger than the eleven-dimensional Planck length and because of this, the above ratio generates a big number. The ansatz that we will make is to consider that the internal metric is of order one in this parameter. If we restrict ourselves to the first few orders of this perturbative expansion we can exclude the contribution that comes from the above mentioned unknown terms. Such a large ``radius'' expansion for the case of Calabi-Yau internal manifold was previously considered in \cite{Witten:1987kg} and \cite{Becker:2001pm}.

The paper is organized as follows. In section \ref{effective-action} we introduce for completeness the low energy effective action of ${\cal M}$-theory with all the known leading quantum correction terms. Also we carefully define the quartic polynomials in the Riemann tensor, that enter in the definition of the quantum correction terms. In section \ref{eq-motion} we analyze perturbatively the equations of motion and we derive conditions that have to be satisfied by the internal background flux in order to have a valid solution. Also at the end of section \ref{eq-motion} we argue that in general the internal manifold losses its Ricci flatness once the quantum correction terms are taken into account. However we show that the manifold remains Ricci flat if a certain condition is satisfied by the warp factors. In section \ref{quartic-polynomials} we discuss some of the properties of the quartic polynomials $E_8$, $J_0$ and $X_8$. These properties are used throughout our analysis presented in section \ref{eq-motion} and we have considered it is useful to have them listed in a separate section. In particular in sub-section \ref{j0-computation} we prove that $J_0$ vanishes on a $Spin(7)$ background and we also derive a compact expression for the first variation of $J_0$ with respect to the internal metric. At the end of sub-section \ref{j0-computation} we compute an elegant formula for the trace of the first variation of $J_0$. In section \ref{susy-solutions} we identify the subset of supersymmetric solutions by analyzing the supersymmetry conditions that emerge from the superpotential. Our brief concluding remarks are shown in section \ref{conclusions}. In appendices we provide the conventions used in this paper as well as some useful identities and small derivations of results which were used in different sections of the paper.


\section{The Low Energy Effective Action}
\label{effective-action}

\setcounter{equation}{0}
\renewcommand{\theequation}{\thesection.\arabic{equation}}


For completeness we introduce in this section the bosonic truncation of the eleven-dimensional supergravity action along with its known correction terms. The effective action for ${\cal M}\,$-theory has the following structure
\begin{equation}
\label{start-point}
S=S_0+S_1 + \ldots \,.
\end{equation}
In the above expression $S_0$ represents the bosonic truncation of eleven-dimensional supergravity \cite{Cremmer:1978km} and $S_1$ represents the leading quantum corrections term. $S_0$ is of order $\kappa_{11}^{-2}$, $S_1$ is of order $\kappa_{11}^{-2/3}$ and the ellipsis denote higher order terms in $\kappa_{11}$. The explicit expressions of $S_0$ and $S_1$ are
\begin{eqnarray}
\label{zero}
S_0 &=&
{1 \over 2\kappa_{11}^2} \int_{M_{11}} d^{11}x\sqrt{-g}R - {1 \over 4\kappa_{11}^2}\int_{M_{11}}
\left(F \wedge \star F + {1 \over 3} C \wedge F \wedge F \right)\,, \\
\label{pppS_1}
S_1 & = & -T_2 \int_{M_{11}} {C \wedge X_8} + b_1 T_2 \int_{M_{11}} {d^{11}x \sqrt{-g} \left( J_0 - {1 \over 2}E_8 \right)} + \ldots \,,
\end{eqnarray}
where $F=dC$ is the four-form field strength of the three-form potential $C$ and $b_1$ is a constant
\begin{equation}
b_1 = {1 \over (2 \pi)^4 3^2 2^{13}} \,.
\end{equation}
$T_2$ is the membrane tension and it is related to the eleven-dimensional gravitational coupling constant by
\begin{equation}
T_2 = \left( \frac{2 \pi^2}{\kappa_{11}^2} \right)^{1/3} \,.
\end{equation}
The components of the eight-form $X_8$ are quartic polynomials of the eleven-dimensional Riemann tensor
\begin{equation}
\label{x8}
X_8(M_{11}) = {1 \over 192 \,(2 \pi)^4} \Big[ {\rm Tr} \, {\cal R}^4 - {1 \over 4} \, ({\rm Tr}\,  {\cal R}^2)^2 \Big] \,,
\end{equation}
where ${\cal R}_{ij} = {1 \over 2} R_{ijkl} \, e^k \wedge e^l$ is the curvature two-form written in some orthonormal frame $e^i$. Furthermore, $E_8$ and $J_0$ are also quartic polynomials of the eleven-dimensional Riemann tensor and have the following expressions
\begin{eqnarray}
\label{pppE_8}
E_8(M_{11}) &=& - \, {1 \over 3!} \, \delta^{ABCM_1 N_1 \ldots M_4 N_4}_{ABCM_1' N_1' \ldots M_4' N_4'} \, {R^{M_1' N_1'}}_{M_1 N_1} \ldots {R^{M_4' N_4'}}_{M_4 N_4} \,, \\
\label{pppJ_0}
J_0(M_{11}) &=& t^{M_1 N_1 \ldots M_4 N_4} \, t_{M_1' N_1' \ldots M_4' N_4'} \, {R^{M_1' N_1'}}_{M_1 N_1} \ldots {R^{M_4' N_4'}}_{M_4 N_4} + {1 \over 4} E_8(M_{11})\,.
\end{eqnarray}
The tensor $t$ is defined by its contraction with some antisymmetric tensor $A$
\begin{equation}
t^{M_1 \ldots M_8} A_{M_1 M_2} \ldots A_{M_7 M_8}
  = 24 {\rm tr} A^4 - 6 ({\rm tr} A^2)^2 \,.
\end{equation}
More details regarding $E_8$, $J_0$ and $X_8$ can be found for example in \cite{Tseytlin:2000sf}.


\section{The Equations of Motion}
\label{eq-motion}

\setcounter{equation}{0}
\renewcommand{\theequation}{\thesection.\arabic{equation}}


In this section we perform a perturbative analysis of the equations of motion and we derive the conditions that the internal flux has to satisfy in order to have a valid solution. We conclude this section with a discussion about the way the internal manifold gets deformed under the influence of higher order correction terms.

The equation of motion which follows from the variation of action (\ref{start-point}) with respect to the metric is
\begin{equation}
\label{eom-g}
R_{MN}(M_{11}) - {1 \over 2} g_{MN} R(M_{11}) - {1 \over 12} T_{MN} = - \beta {1\over \sqrt{-g}} {\delta \over \delta g^{MN}} \left[ \sqrt{-g} (J_0-{1\over 2}E_8) \right] \,,
\end{equation}
where $T_{MN}$ is the energy momentum tensor of $F$ given by
\begin{equation}
\label{en-mom-tensor}
T_{MN}=F_{MPQR} \, {F_N}^{PQR}-{1 \over 8} \, g_{MN} \, F_{PQRS} \, F^{PQRS} \,,
\end{equation}
and we have set $\beta=2 \kappa_{11}^2 b_1 T_2$. We have listed in appendix \ref{scaling} the expressions for the external and internal energy-momentum tensor. Also in the above mentioned appendix we provide the results obtained for the external and internal components of the term in right hand side of the Einstein equation (\ref{eom-g}).

Without sources the field strength obeys the Bianchi identity
\begin{equation}
dF=0 \,,
\end{equation}
and the equation of motion
\begin{equation}
\label{eom-F}
d*F={1\over 2}F \wedge F+ {{\beta} \over {b_1}}X_8 \,.
\end{equation}
The metric ansatz is a warped product
\begin{eqnarray}
ds^2 &=& g_{MN} \, dX^M \, dX^N = \widetilde{\eta}_{\mu \nu }(x,y) \, dx^{\mu} dx^{\nu} + \widetilde{g}_{mn}(y) \, dy^m dy^n  \nonumber \\
\label{full-metric}
&=& e^{2A(y)} \, \eta_{\mu \nu } (x) \, dx^{\mu} dx^{\nu} + e^{2B(y)} \, g_{mn}(y) \, dy^m dy^n \,,
\end{eqnarray}
where $\eta_{\mu\nu}$ describes a three-dimensional external space $M_3$ and $g_{mn}$ is a $Spin(7)$ holonomy metric of a compact manifold $M_8$. As usual the big Latin indices $M$, $N$ take values between $0$ and $10$, the Greek indices $\mu$, $\nu$ take values between $0$ and $2$ and small Latin indices $m$, $n$ take values between $3$ and $10$. Also, $X^M$ refers to the coordinates on the whole eleven-dimensional manifold $M_{11}$, $x^\mu$ are the coordinates on $M_3$ and $y^m$ are the coordinates on $M_8$. We want to note that $M_{11}$ is the direct product between $M_3$ and $M_8$ only in the leading order approximation.

In what follows we introduce a dimensionless parameter ``$t$'' defined as the square of the ratio between $l_8$, the characteristic size of the internal manifold $M_8$, and $l_{11}$ which denotes the eleven-dimensional Plank length
\begin{equation}
\label{t-def}
t=\left( {l_8 \over l_{11}} \right)^2 \gg 1 \,,
\end{equation}
where $l_8$ is given by
\begin{equation}
(l_8)^8 = \int_{M_8} d^8y \, \sqrt{g} \,,
\end{equation}
and we have considered that $l_8 \gg l_{11}$. We will suppose that all the fields of the theory have a series expansion in ``$t$'' and we will analyze the equations of motion order by order in the ``$t$'' perturbative expansion. We consider that the metric of the internal compact space $M_8$ has the following series expansion in ``$t$''
\begin{equation}
\label{metric-exp}
g_{mn}=t \, [g^{(1)}]_{mn} + [g^{(0)}]_{mn} + \ldots \,.
\end{equation}
Thus the inverse metric is
\begin{equation}
g^{mn} = t^{-1} [g^{(1)}]^{mn} + t^{-2} [g^{(2)}]^{mn} + \ldots \,,
\end{equation}
where the expressions of the above expansion coefficients are derived in appendix \ref{scaling}. It is obvious now that the Riemann tensor, the Ricci tensor and the Ricci scalar of the internal manifold $M_8$ will have a series expansion in ``$t$'' of the form
\begin{eqnarray}
{R^a}_{ ~mbn}(M_8) &=& {[R^{(0)}]^a}_{mbn} + t^{-1} {[R^{(1)}]^a}_{mbn} + \ldots \,, \\
R_{mn}(M_8) & = & [R^{(0)}]_{mn} + t^{-1} [R^{(1)}]_{mn} + \ldots\,, \\
R(M_8) &=& t^{-1} R^{(1)} +t^{-2} R^{(2)} + \ldots \,,
\end{eqnarray}
where the coefficients in the above expansions can be determined in terms of the expansion coefficients of $g_{mn}$ and $g^{mn}$ and their derivatives. It is not so obvious at this stage of computation that the right ansatz for the warp factors is
\begin{equation}
\label{warp-exp}
X=t^{-3} \, X^{(3)} + \ldots \quad \Rightarrow \quad e^{X}=1+t^{-3} \, X^{(3)}+\ldots \,,
\end{equation}
with $X=A,B$. The motivation for this ansatz comes from the fact that the internal Einstein equation receives contributions from the quantum correction terms in the $t^{-3}$ order of perturbation theory. It is natural to suppose that the effect of warping appears at the same order in the equations of motion to compensate for this extra contribution.

The Poincare invariance restricts the form of the background flux $F$ to the following structure
\begin{equation}
\label{flux-form}
F = F_1 + F_2 \,,
\end{equation}
where $F_1$ is the external part of the flux
\begin{equation}
F_1 = {1 \over {3!}} \, \epsilon_{\mu \nu \rho} \, [\nabla_m f(y)] \, dx^\mu \wedge dx^\nu \wedge dx^\rho \wedge dy^m \,,
\end{equation}
and $F_2$ is the internal background flux
\begin{equation}
F_2 = {1 \over {4!}} \, F_{mnpq}(y) \, dy^m \wedge dy^n \wedge dy^p \wedge dy^q \,.
\end{equation}
We also expand $f$ and $F$ in a power series of $t$
\begin{equation}
f=f^{(0)}+t^{-1}f^{(1)}+\ldots \,,
\end{equation}
and
\begin{equation}
F=F^{(0)}+t^{-1}F^{(1)}+\ldots \,.
\end{equation}

Taking into account that the three-dimensional space described by $\eta_{\mu \nu}$ is not at all influenced by the size of the eight-dimensional manifold $M_8$ described by $g_{mn}$, all the quantities that emerge from the metric $\eta_{\mu \nu}$ are of order zero in an expansion in ``$t$'', in other words all these quantities are independent of the scale ``$t$''. The external manifold suffers no change due to the deformations of the internal manifold and $\eta_{\mu \nu}$ generates the same equations of motion as in the absence of fluxes and without the quantum correction terms. The zeroth order of the external component of equation (\ref{eom-g}) reads
\begin{equation}
R_{\mu\nu}(M_3) - {1 \over 2} \eta_{\mu\nu} R(M_3) = 0 \,,
\end{equation}
therefore
\begin{eqnarray}
R_{\mu\nu}(M_3) &=& 0 \,, \\
R(M_3) &=& 0 \,.
\end{eqnarray}

A careful analysis of the internal and external Einstein equations to orders no higher than $t^{-2}$ and $t^{-3}$ respectively reveals that the internal manifold remains Ricci flat to the $t^{-2}$ order in perturbation theory
\begin{equation}
\label{Ricci-t2}
R^{(0)}_{mn} = R^{(1)}_{mn}=R^{(2)}_{mn}=0 \,,
\end{equation}
and the Ricci scalar vanishes to the $t^{-3}$ order
\begin{equation}
R^{(1)} = R^{(2)} = R^{(3)} = 0 \,.
\end{equation}
These results are natural because we expect to observe deformations of the internal manifold starting at the $t^{-3}$ order since the quantum correction terms are of this order of magnitude in an expansion in ``$t$'' and in addition the warp factors were chosen to be of the same order of magnitude. As a matter of fact, to order $t^{-2}$ even the warping has no effect and the eleven-dimensional manifold is a direct product between $M_3$ and $M_8$.

We can also derive from the equation of motion (\ref{eom-F}) that the covariant derivative of the external flux vanishes to order $t^{-2}$
\begin{equation}
\nabla_m f^{(0)} = \nabla_m f^{(1)} = \nabla_m f^{(2)} = 0 \,.
\end{equation}
Collecting these facts we are left with the following field decomposition for $R_{mn}(M_8)$, $R(M_8)$ and $\nabla_m f$
\begin{eqnarray}
R(M_8) &=& t^{-4} R^{(4)} + \ldots \,, \\
R_{mn}(M_8) &=& t^{-3} \, R^{(3)}_{mn} + t^{-4} \, R^{(4)}_{mn} + \ldots \,, \\
\nabla_m f &=& t^{-3} \nabla_m f^{(3)} + t^{-4} \nabla_m f^{(4)} + \ldots \,.
\end{eqnarray}

To order $t^{-4}$ the external component of the equation of motion (\ref{eom-g}) has the following form
\begin{equation}
\label{external-t3}
R^{(4)} - 4 \triangle^{(1)} A^{(3)} - 14 \triangle^{(1)} B^{(3)} - {1\over 48} \left[F^{(0)} \right]^2+{\beta \over 2} E_8^{(4)}(M_8) =0 \,,
\end{equation}
where we have defined
\begin{equation}
\label{laplace-p}
\triangle^{(1)} = [g^{(1)}]^{mn} \, \nabla_m \nabla_n \,,
\end{equation}
and
\begin{equation}
\left[F^{(0)} \right]^2 = [g^{(1)}]^{aa'} \, [g^{(1)}]^{bb'} \, [g^{(1)}]^{mm'} \, [g^{(1)}]^{nn'} \, F^{(0)}_{abmn} \, F^{(0)}_{a'b'm'n'} \,.
\end{equation}
We note that the right hand side of (\ref{eom-g}) has been evaluated on the un-warped background because to this order the warping is not felt by that term. To order $t^{-4}$ in perturbation theory the trace of the internal Einstein equation has the following form
\begin{equation}
\label{internal-trace-t4}
3[R^{(4)} - 7 \triangle^{(1)} A^{(3)} -14 \triangle^{(1)} B^{(3)}] = 2^{15} \, \beta \, \triangle^{(1)} \left[4 \, E_6^{(3)}(M_8)  +  \Omega^{(2)} \cdot z^{(5)} \right] \,.
\end{equation}
Eliminating the $R^{(4)}$ term from equations (\ref{external-t3}) and (\ref{internal-trace-t4}) we obtain an equation for the warp factor $A^{(3)}$ and the internal flux $F^{(0)}$
\begin{equation}
\label{relation-01}
3 \triangle^{(1)} A^{(3)}  - {1\over 48} \left[F^{(0)} \right]^2+{\beta \over 2} E_8^{(4)}(M_8) - 2^{15} \, \beta \, \triangle^{(1)} \left[4 \, E_6^{(3)}(M_8)  +  \Omega^{(2)} \cdot z^{(5)} \right] = 0 \,.
\end{equation}
The equation of motion  for the external flux at the order $t^{-4}$ is \cite{Becker:2001pm}
\begin{equation}
\label{relation-02}
\triangle^{(1)} f^{(3)}- {1\over 48} F^{(0)} \star^{(1)} F^{(0)}+ {\beta \over 2} E_8^{(4)}(M_8) = 0 \,,
\end{equation}
where the Hodge $\star^{(1)}$ operation is performed with respect to the leading order term $[g^{(1)}]_{mn}$ of the internal metric. If we subtract (\ref{relation-02}) from (\ref{relation-01}) and integrate\footnote{The integration is performed on a manifold which we have denoted $M'_8$, whose metric is $[g^{(1)}]_{mn}$. In some sense we can think of $[g^{(1)}]_{mn}$ as being the undeformed $Spin(7)$ holonomy metric and the next order term $[g^{(0)}]_{mn}$ being the deformation from the exceptional holonomy metric. Hence $M'_8$ can be thought as the undeformed $Spin(7)$ holonomy manifold. We also want to note that $E_8^{(4)}(M_8)$ is the Euler integrant of $M'_8$.} the resulting expression we obtain that $F^{(0)}$ is self dual with respect to $\star^{(1)}$ operation
\begin{equation}
\label{self-duality}
\star^{(1)} F^{(0)} = F^{(0)} \,,
\end{equation}
and it satisfies
\begin{equation}
\label{tadpole-anomaly}
\frac{1}{4 \kappa^2_{11}} \, \int_{M'_8} F^{(0)} \wedge F^{(0)} = \frac{T_2}{24} \, \chi_8\,' \,,
\end{equation}
where $\chi_8\,'$ is the Euler character of $M'_8$. The last relation is obtained from integrating out the equation (\ref{relation-02}) and considering that the internal flux is self dual. The condition (\ref{tadpole-anomaly}) is nothing else but the perturbative leading order of the global tadpole anomaly relation\footnote{We remind the reader that we have not considered space-filling membranes in our calculations.} that the internal flux has to obey when compactifications of ${\cal M}$-theory on eight-dimensional manifolds are taken into consideration \cite{Sethi:1996es}.

The difference between equations (\ref{relation-02}) and (\ref{relation-01}) together with the self duality condition (\ref{self-duality}) of the internal flux produces an equation which relates the warp factor $A$ to the external flux
\begin{equation}
\label{warp-eflux}
\triangle^{(1)} \, \left\{ f^{(3)} - 3 A^{(3)} + 2^{15} \, \beta \, \left[4 \, E_6^{(3)}(M_8)  +  \Omega^{(2)} \cdot z^{(5)} \right] \right\}= 0 \,.
\end{equation}
Also the self-duality of the internal flux implies the vanishing of the following expression
\begin{equation}
[F^{(0)}]_{mabc} \, {[F^{(0)}]_n}^{abc} - {1 \over 8} \, [g^{(1)}]_{mn} \, [F^{(0)}]_{abcd} \, [F^{(0)}]^{abcd} = 0 \,,
\end{equation}
where the details of the derivation are provided in the appendix B of \cite{deWit:1978sh}. Therefore we are left with the following form for the internal Einstein equation to the order $t^{-3}$ in perturbation theory
\begin{equation}
\label{internal-t3}
R^{(3)}_{mn} - {1 \over 2} \, g^{(1)}_{mn} \, R^{(4)} + 3 [ \, g^{(1)}_{mn} \triangle^{(1)} - \nabla_m \nabla_n] \, C^{(3)} + \beta \, \left(\frac{\delta Y}{\delta g^{mn}}\right)^{(3)} = 0 \,,
\end{equation}
where ${\delta Y}/{\delta g^{mn}}$ and its trace are computed in section \ref{j0-computation} and we have denoted
\begin{equation}
C=A+2B \,.
\end{equation}
The internal manifold remains Ricci-flat only under a very specific condition. To determine this condition we replace in (\ref{internal-t3}) the expression for the perturbative coefficient of the Ricci scalar $R^{(4)}$ obtained from (\ref{internal-trace-t4})
\begin{equation}
R^{(4)} = 7 \triangle^{(1)} C^{(3)} - \, \frac{\beta}{3} \, \left(g^{ab} \frac{\delta Y}{\delta g^{ab}}\right)^{(4)} \,,
\end{equation}
and we recast (\ref{internal-t3}) in the following form
\begin{equation}
\label{i-t3}
R^{(3)}_{mn} + {1 \over 6} \, g^{(1)}_{mn} \, \left[ \beta \, \left(g^{ab} \frac{\delta Y}{\delta g^{ab}}\right)^{(4)} - 3 \triangle^{(1)} C^{(3)} \right] + \left[ \beta \, \left(\frac{\delta Y}{\delta g^{mn}}\right)^{(3)} - 3 \nabla_m \nabla_n \, C^{(3)} \right] = 0 \,.
\end{equation}
Now it is easy to see that Ricci flatness to this order in perturbation theory requires that
\begin{equation}
\label{R-flat-condition}
\nabla_m \nabla_n C^{(3)} = \frac{\beta}{3} \left(\frac{\delta Y}{\delta g^{mn}}\right)^{(3)} \,,
\end{equation}
which is a strong constraint on the warp factors. We can conclude that the internal manifold gets modified at the $t^{-3}$ order in perturbation theory in the sense that in general it looses its Ricci flatness unless very restrictive constraints are imposed to the warp factors.


\section{Some Properties of the Quartic Polynomials}
\label{quartic-polynomials}


\setcounter{subsection}{0}


\setcounter{equation}{0}
\renewcommand{\theequation}{\thesection.\arabic{equation}}


In this section we look at some of the properties related to the quartic polynomials in the Riemann tensor which appear in the low energy effective action of ${\cal M}$-theory. More precisely we will derive several relations obeyed by the polynomials which appear in the definition (\ref{pppS_1}) of $S_1$. There are three different subsections, one for each of the polynomials $E_8$, $J_0$ and $X_8$. We want to emphasize that all the properties of these polynomials are computed on an undeformed background, i.e. our background is a direct product $M_3 \times M_8$ with $M_3$ being maximally symmetric and $M_8$ being a $Spin(7)$ holonomy manifold. Obviously the warping and the deformation of the background will correct all the relations derived in the following subsections but these corrections are of a higher order than $t^{-4}$ and we can neglect them as our analysis stops at this order in perturbation theory.


\subsection{Properties of $E_8$ Polynomial}
\label{e8-computation}

\setcounter{equation}{0}
\renewcommand{\theequation}{\thesubsection.\arabic{equation}}


Let us focus now on the properties of the quartic polynomial $E_8$ defined in (\ref{pppE_8}) for an eleven-dimensional manifold. We can generalize its definition by introducing a polynomial $E_n({M_D})$ for any even $n$ and any $D\,$-dimensional manifold $M_D$ ($n \le D$) as follows
\begin{equation}
\label{generalized-euler}
E_n(M_D) = \pm \, \delta^{M_1 \ldots M_n}_{K_1 \ldots K_n} \, {R^{K_1 K_2}}_{M_1 M_2} \ldots {R^{K_{n-1} K_n}}_{M_{n-1} M_n} \,,
\end{equation}
where the indices take values from $0$ to $D-1$ and the ``$+$'' corresponds to the Euclidean signature and the ``$-$'' corresponds to the Lorentzian signature. As we have mentioned at the beginning of section \ref{quartic-polynomials}, $E_8$ is computed on a direct product manifold $M_{11}=M_3 \times M_8$ therefore we have \cite{Haack:2001jz}
\begin{equation}
\label{e8-property}
E_8(M_3 \times M_8) = - E_8(M_8) - 8 R(M_3) E_6(M_8) = - E_8(M_8) \,,
\end{equation}
where $R(M_3)$ is the Ricci scalar for the external manifold which is zero in our case. If $n=D$ in formula (\ref{generalized-euler}) then $E_n(M_n)$ is proportional to the Euler integrand of $M_n$. In particular for $E_8(M_8)$ we have that
\begin{equation}
\label{e8-euler}
\int_{M_8} E_8(M_8) \sqrt{g} \, d^8y = \frac{\chi_8}{12\,b_1} \,,
\end{equation}
where $\chi_8$ is the Euler character of $M_8$. If the manifold $M_8$ has a nowhere-vanishing spinor, $E_8(M_8)$ and $X_8(M_8)$ are related in the sense that their integrals over $M_8$ are proportional to the Euler characteristic of $M_8$. The details of this correspondence are provided in section \ref{x8-computation}. The variation of $E_8(M_8)$ with respect to the internal metric can be derived using the definition (\ref{generalized-euler}) or much easier from (\ref{e8-euler})
\begin{equation}
\label{e8-variation}
\frac{\delta E_8(M_8)}{\delta g_{mn}} =  - {1 \over 2} \, g^{mn} \, E_8(M_8) \,,
\end{equation}
therefore the trace of the variation is
\begin{equation}
\label{e8-trace}
g_{mn} \frac{\delta E_8(M_8)}{\delta g_{mn}} = - 4 E_8(M_8) \,.
\end{equation}
We want to note that the variation of $E_8$ given in (\ref{e8-variation}) is of order $t^{-5}$ whereas its trace (\ref{e8-trace}) is of order $t^{-4}$. Finally, for further reference, we provide the perturbative expansion for $E_8(M_8)$ and $E_6(M_8)$
\begin{eqnarray}
\label{e8-expansion}
E_8(M_8) &=& t^{-4} \, {E_8}^{(4)}(M_8) + \ldots \,, \\
\label{e6-expansion}
E_6(M_8) &=& t^{-3} \, {E_6}^{(3)}(M_8) + \ldots \,.
\end{eqnarray}
%


\subsection{Properties of $J_0$ Polynomial}
\label{j0-computation}

\setcounter{equation}{0}
\renewcommand{\theequation}{\thesubsection.\arabic{equation}}


In this subsection we look closely at the properties of the quartic polynomial $J_0$ defined in (\ref{pppJ_0}). We particularize the background to be $Spin(7)$ holonomy compact manifold and we compute the value of $J_0$ integral on such a background. We will also calculate the first variation of $J_0$ with respect to the internal metric and the trace of its first variation.

As we anticipated in our previous work \cite{Becker:2003wb}, for a $Spin(7)$ holonomy manifold the integral of the quartic polynomial $J_0$ vanishes. Bellow we provide the detailed proof of this statement. The essential fact that constitutes the basis of the demonstration is the existence of the covariantly constant spinor on a compact manifold which has $Spin(7)$ holonomy.

The quartic polynomial $J_0$ can be expressed as a sum of an internal and an external polynomial \cite{Haack:2001jz}. Furthermore, these polynomials can be written only in terms of the internal and external Weyl tensors \cite{Banks:1998nr, Gubser:1998nz}. Since the Weyl tensor vanishes in three dimensions we are left only with the contribution from the internal polynomial
\begin{equation}
\label{j0-integral-11d}
\int_{M_{11}} J_0(M_{11}) \, \sqrt{-g} \, d^{11}x = \int_{M_8} J_0(M_{8}) \, \sqrt{g} \, d^8y \,.
\end{equation}
Because the internal manifold has a nowhere-vanishing spinor, the integral of the remaining internal part can be replaced by the kinematic factor which appears in the four-point scattering amplitude for gravitons, as explained in \cite{Isham:1988jb}
\begin{equation}
\int_{M_8} J_0(M_8) \, \sqrt{g} \, d^8y = \int_{M_8} Y \, \sqrt{g} \, d^8y\,,
\end{equation}
where we have denoted the kinematic factor with $Y$. As a matter of fact $J_0$ represents the covariant generalization of $Y$ and the modifications of the equations of motion are given in terms of $Y$ and its variation with respect to the internal metric. This kinematic factor was written in \cite{Gross:1986iv} as an integral over $SO(8)$ chiral spinors\footnote{The eight-rank tensor $``t''$ that appears in \cite{Gross:1986iv} is different from our convention.}
\begin{equation}
\label{so8-integral}
Y = \int d \psi_L\, d \psi_R \exp(R_{abmn}\, \bar\psi_L\, \Gamma^{ab}\, \psi_L\, \bar\psi_R\, \Gamma^{mn}\, \psi_R)\,,
\end{equation}
where (\ref{so8-integral}) is evaluated using the rules of Berezin integration. As argued in \cite{Gross:1986iv}, $Y$ is zero for Ricci-flat and K\"{a}hler manifolds, but for general Ricci-flat manifolds it does not necessarily have to vanish. In our case, the ${\bf 8}_s$ of $SO(8)$ decomposes under $Spin(7)$ as ${\bf 7} \oplus {\bf 1}$. The singlet in this decomposition corresponds to the Killing spinor $\eta$ of the $Spin(7)$ manifold. If the holonomy group of the eight-dimensional manifold is $Spin(7)$ and not some proper subgroup, then the covariantly constant spinor $\eta$ is the only zero mode of the Dirac operator, as proved in \cite{Joyce}. Moreover, the parallel spinor obeys the integrability condition (e.g. see \cite{Gibbons:1990er})
\begin{equation}
R_{abmn} \Gamma^{mn} \eta = 0 \,,
\end{equation}
therefore the integrand of (\ref{so8-integral}) does not depend on the Killing spinor $\eta$ and implies the vanishing of $Y$ for $M_8$ with $Spin(7)$ holonomy
\begin{equation}
Y = 0 \quad \mathrm{for} \quad \mathrm{Hol}[g(M_8)] = Spin(7) \,,
\end{equation}
which implies the vanishing of the integral (\ref{j0-integral-11d}). It has been shown in \cite{Lu:2003ze} that $Y$ vanishes in the $G_2$ holonomy case as well. The Calabi-Yau case is another example where the polynomial $Y$ vanishes \cite{Gross:1986iv}. The fact that the manifold is Ricci-flat and K\"{a}hler ensures the existence of the covariantly constant spinors, which is sufficient to imply $Y=0$ as explained in \cite{Freeman:1986br}. We conclude that the integral of $J_0$ vanishes if the internal manifold admits at least one covariantly constant spinor, in particular it vanishes for an internal manifold which has $Spin(7)$ holonomy.

In what follows we will derive the first variation of $Y$ with respect to the internal metric. One can use (\ref{so8-integral}) to compute the variation of $Y$ and the following result is obtained \cite{Lu:2003ze}
\begin{eqnarray}
\delta Y &=& 4 \, \epsilon^{\alpha_1 \cdots \alpha_8} \, \epsilon^{\beta_1 \cdots \beta_8} \, (\Gamma^{i_1 i_2})_{\alpha_1 \alpha_2} \cdots (\Gamma^{i_7 i_8})_{\alpha_7 \alpha_8} \, (\Gamma^{j_1 j_2})_{\beta_1 \beta_2} \cdots (\Gamma^{j_7 j_8})_{\beta_7 \beta_8}\, \times \nonumber \\
\label{Y-variation-1}
&& R_{i_1 i_2 j_1 j_2} \, R_{i_3 i_4 j_3 j_4} \, R_{i_5 i_6 j_5 j_6} \, \delta R_{i_7 i_8 j_7 j_8} \,.
\end{eqnarray}
Because the internal manifold has a nowhere vanishing spinor we can transform from the spinorial representation to the vector representation ${\bf 8_v}$ of $SO(8)$. From \cite{Gibbons:1990er} we have the following relation between these representations
\begin{equation}
V^a = -i (\overline{\eta} \Gamma^a)_\alpha \psi^\alpha \,,
\end{equation}
where $\eta$ is the unit Killing spinor. After performing the change of representation in  (\ref{Y-variation-1}) and using the identity (\ref{id3}) and relation (\ref{Riemann-variation}) we obtain
\begin{equation}
\delta Y = - 2^{15} \, z^{k_7 k_8 m_7 m_8} \, {{\overline{\Omega}}^{\, i_7 i_8}}_{k_7 k_8} \, {\overline{\Omega}^{\, j_7 j_8}}_{m_7 m_8} \, \nabla_{i_7} \nabla_{j_7} \delta g_{i_8 j_8} \,,
\end{equation}
where we have introduced
\begin{equation}
\label{omega-bar}
{\overline{\Omega}^{\,ab}}_{mn} = {\Omega^{ab}}_{mn} + \delta^{ab}_{mn} \,,
\end{equation}
$\Omega$ being the Cayley calibration of the $Spin(7)$ holonomy manifold $M_8$. To provide a perturbative expansion for $\Omega$ we have to remember that the volume ${\cal V}_{M_8}$ of the internal manifold $M_8$ can be expressed in terms of the Cayley calibration
\begin{equation}
\int \Omega \wedge \star \Omega = 14 {\cal V}_{M_8} \,,
\end{equation}
hence it is very easy to realize that the Cayley calibration perturbative expansion is
\begin{equation}
\label{Cayley-expansion}
\Omega_{mnpr} = t^2 \, \Omega_{mnpr}^{(2)} + t \, \Omega_{mnpr}^{(1)} + \ldots \,.
\end{equation}
The polynomial $z^{k_7 k_8 m_7 m_8}$ is cubic in the eight-dimensional Riemann tensor and it is defined by
\begin{equation}
\label{z-tensor}
z^{k_7 k_8 m_7 m_8} = |g|^{-1} \, \epsilon^{a_1 \cdots a_6 k_7 k_8}\, \epsilon^{b_1 \cdots b_6 m_7 m_8} \, R_{a_1 a_2 b_1 b_2} \, R_{a_3 a_4 b_3 b_4} \, R_{a_5 a_6 b_5 b_6} \,.
\end{equation}
It is obvious that the perturbative expansion of $z^{mnpr}$ has the following form
\begin{equation}
z^{mnpr} = t^{-5} \, {[z^{(5)}]}^{mnpr} + \ldots \,.
\end{equation}

Finally we determine the expression of the first variation of $Y$ with respect to the internal metric
\begin{equation}
\label{Y-variation}
\frac{\delta Y}{\delta g_{i_8 j_8}} = - 2^{15} \, {{\overline{\Omega}}^{\,i_7 i_8}}_{k_7 k_8} \, {\overline{\Omega}^{\,j_7 j_8}}_{m_7 m_8} \nabla_{i_7} \nabla_{j_7} z^{k_7 k_8 m_7 m_8} \,,
\end{equation}
which contributes at the internal Einstein equation. It is obvious that the leading order of (\ref{Y-variation}) is $t^{-5}$, i.e. the leading order of $z^{mnpr}$. However the term $\delta Y / \delta g^{ij}$ which appears in the equation of motion (\ref{eom-g}) is of order $t^{-3}$. In other words, $t^{-3}$ is the order at which the equations of motion receive contributions from the quantum correction terms. As we have explained in section \ref{eq-motion}, it is natural to suppose that the warping effects are visible to the same order in the perturbation theory and this is why we have considered the ansatz (\ref{warp-exp}).

In addition we also need the trace of (\ref{Y-variation}) with respect to the internal metric. We provide in what follows the main steps of the derivation. We begin the computation by using the definition (\ref{omega-bar}) for ${\overline{\Omega}}$ and we obtain that
\begin{eqnarray}
g_{i_8 j_8} \frac{\delta Y}{\delta g_{i_8 j_8}} &=& - 2^{15} \, [ \, g_{i_8 j_8} {\Omega^{i_7 i_8}}_{k_7 k_8} \, {\Omega^{j_7 j_8}}_{m_7 m_8} \nabla_{i_7} \nabla_{j_7} z^{k_7 k_8 m_7 m_8} \nonumber \\
&+& g_{i_8 j_8} \, {\Omega^{i_7 i_8}}_{k_7 k_8} \, \delta^{j_7 j_8}_{m_7 m_8} \nabla_{i_7} \nabla_{j_7} z^{k_7 k_8 m_7 m_8} \nonumber \\
&+& g_{i_8 j_8} \, \delta^{i_7 i_8}_{k_7 k_8} \, {\Omega^{j_7 j_8}}_{m_7 m_8} \nabla_{i_7} \nabla_{j_7} z^{k_7 k_8 m_7 m_8} \nonumber \\
\label{first-step}
&+& 4 \nabla^a \nabla_b \left( {z_{am}}^{bm} \right) \, ] \,.
\end{eqnarray}
We denote the first, the second and the third terms in the square parentheses of (\ref{first-step}) with $T_1$, $T_2$ and $T_3$ respectively. Using (\ref{omega-contraction-1}), $T_1$ can be rewritten as
\begin{eqnarray}
T_1 &=& 2 \triangle \left({z_{mn}}^{mn} \right) + \triangle \left( \Omega \cdot z \right) \nonumber \\
&& - 2  \left( \nabla_a \nabla^b +\nabla^b \nabla_a \right) \left( \Omega \cdot z \right) - 4 \nabla^a \nabla_b \left( {z_{am}}^{bm} \right) \,,
\end{eqnarray}
where $\triangle = \nabla_a \nabla^a$ and $\Omega \cdot z$ is a short notation for the full contraction between the Cayley calibration $\Omega$ and the $z$ polynomial. The sum of the second and the third terms in (\ref{first-step}) can be rewritten as
\begin{equation}
T_2 + T_3 = 2  \left( \nabla_a \nabla^b +\nabla^b \nabla_a \right) \left( \Omega \cdot z \right) \,.
\end{equation}
Therefore we obtain an elegant and compact expression for the trace of (\ref{Y-variation})
\begin{equation}
\label{trace-j0}
g_{mn} \frac{\delta Y}{\delta g_{mn}} = - 2^{15} \, \triangle \left(2{z_{mn}}^{mn}  + \Omega \cdot z \right) \,.
\end{equation}
With the observation that
\begin{equation}
{z_{mn}}^{mn} = 2 \, E_6(M_8) \,,
\end{equation}
the result (\ref{trace-j0}) can be expressed as
\begin{equation}
\label{Y-trace}
g_{mn} \frac{\delta Y}{\delta g_{mn}} =  - 2^{15} \, \triangle \left[4 \, E_6(M_8)  +  \Omega \cdot z \right] \,,
\end{equation}
where $E_6(M_8)$ is given by (\ref{generalized-euler}) for $n=6$ and $D=8$. It is very interesting to note the close resemblance of formula (\ref{Y-trace}) with the corresponding one for Calabi-Yau manifolds \cite{Becker:2001pm}. We also want to emphasize that the shift of the Cayley calibration toward $\overline{\Omega}$ is exactly what is needed in order to obtain the simple form of the trace given in (\ref{Y-trace}). A simple analysis of formula (\ref{Y-trace}) reveals that the trace of the first variation of $Y$ is of order $t^{-4}$.
\begin{equation}
g_{mn} \frac{\delta Y}{\delta g_{mn}} =  - 2^{15} \, \triangle^{(1)} \left[4 \, E_6^{(3)}(M_8)  +  \Omega^{(2)} \cdot z^{(5)} \right] \, t^{-4} + \ldots \,,
\end{equation}
where $E_6^{(3)}(M_8)$ was introduced in equation (\ref{e6-expansion}), $\triangle^{(1)}$ was defined in (\ref{laplace-p}) and $\Omega^{(2)}$ is a coefficient which appears in the perturbative expansion (\ref{Cayley-expansion}) of the Cayley calibration.


\subsection{Properties of $X_8$ Polynomial}
\label{x8-computation}

\setcounter{equation}{0}
\renewcommand{\theequation}{\thesubsection.\arabic{equation}}


The integral of $X_8(M_8)$ over an eight-dimensional manifold $M_8$ is related to the Euler characteristic $\chi_8$ of the manifold if $M_8$ admits at least one nowhere vanishing spinor
\begin{equation}
\label{x8-euler}
\int_{M_8} X_8(M_8) = - \, {\chi_8 \over 24} \,.
\end{equation}
In our paper $M_8$ has $Spin(7)$ holonomy, so there is a Killing spinor on $M_8$ and therefore we can use the above property in our derivations.

In what follows we will justify the relation (\ref{x8-euler}). The eight-form $X_8$ is defined by relation (\ref{x8}) and can be expressed in terms of the first two Pontryagin forms $P_1$ and $P_2$
\begin{equation}
X_8 = {1 \over 192} \, [P_1^2-4P_2]\,,
\end{equation}
where
\begin{eqnarray}
P_1 &=& - {1 \over 8 \pi^2} {\rm Tr} \,{\cal R}^2 \,, \\
P_2 &=&  {1 \over 128 \pi^4} \, [({\rm Tr} \, {\cal R}^2)^2 - 2 {\rm Tr} \, {\cal R}^4] \,,
\end{eqnarray}
and ${\cal R}$ is the curvature two-form. The existence of a covariantly constant spinor on a $Spin(7)$ holonomy manifold means that we have a nowhere vanishing spinor field on the eight-dimensional manifold. It has been shown in \cite{Isham:1988jb} that under these circumstances there is a necessary and sufficient condition which relates the Euler class and the first two Pontryagin classes of the manifold
\begin{equation}
e - {1 \over 2} \, P_2 + {1 \over 8} \, P_1^2 = 0 \,,
\end{equation}
where $e$ is the Euler integrand of $M_8$. Hence, the eight-form $X_8$ is proportional to the Euler integrand of $M_8$
\begin{equation}
X_8(M_8) = - {1 \over {24}} \, e(M_8) \,,
\end{equation}
and from here the relation in (\ref{x8-euler}) follows immediately.


\section{The Set of Supersymmetric Solutions}
\label{susy-solutions}

\setcounter{equation}{0}
\renewcommand{\theequation}{\thesection.\arabic{equation}}


We have determined in section \ref{eq-motion} that the internal flux is a self-dual four form. Since the expressions in which the internal flux appears are of order $t^{-4}$ or higher our analysis will be performed on an undeformed and unwarped background as we have previously proceeded in section \ref{quartic-polynomials}. As a consequence in this section the manifold $M_8$ is considered to be a $Spin(7)$ holonomy manifold. Under these circumstances it is very easy to select the supersymmetric solutions. We want to emphasize that our discussion involves the leading order perturbative expansion coefficients of the Cayley calibration $\Omega^{(2)}$ and the one for the internal flux $F^{(0)}$. However, the discussion is not influenced by this fact because the perturbative expansion of the fields does not affect their representation, in other words all the expansion coefficients will belong to the same representation as the original field. Our analysis is based only on the fields representations therefore in what follows we drop the upper index which denote the order in perturbation theory and the discussion will make no distinction between the full quantities $F$ and $\Omega$ and their leading contributions $F^{(0)}$ and $\Omega^{(2)}$, respectively.

The constraints imposed to the internal flux in order to obtain a supersymmetric vacua in three dimensions have been determined in \cite{Becker:2000jc}. Later on in \cite{Acharya:2002vs} these constraints have been derived from certain conditions imposed to the superpotential $W$ whose expression was conjectured in \cite{Gukov:1999gr} to be
\begin{equation}
\label{superpotential}
W=\int_{M_8} F \wedge \Omega \,.
\end{equation}
In \cite{Becker:2002jj} we have shown that (\ref{superpotential}) is indeed the superpotential of the theory and in a consequent paper \cite{Becker:2003wb} we have proved that the scalar potential of the effective three-dimensional theory is
\begin{equation}
\label{scalar-potential}
V[W] = \sum_{A,B=1}^{b^4_{{\bf 35}^-}} {\cal L}^{AB} \, D_AW \, D_BW \,,
\end{equation}
where ${\cal L}^{AB}$ is the inverse matrix of
\begin{equation}
\label{moduli-metric-l2}
{\cal L}_{AB} = {1 \over {\cal V}_{M_8}} \, \int \xi_A \wedge \star \, \xi_B \,,
\end{equation}
and ${\cal V}_{M_8}$ is the volume of $M_8$. We have denoted with $b^4_{{\bf 35}^-}$ the refined Betti number which represents the number of anti-self-dual harmonic four-forms $\xi_A$. The covariant derivative $D_A$ is defined through its action on the Cayley calibration\footnote{For more details the reader can consult \cite{Becker:2003wb}.}
\begin{equation}
D_A \, \Omega =\xi_A \,.
\end{equation}
As we have discussed in \cite{Becker:2003wb} the minimum of the scalar potential (\ref{scalar-potential}) is zero due to its quadratic expression. Therefore an $AdS_3$ solution is excluded and a three-dimensional supersymmetric effective theory is obtained only when the scalar potential vanishes, i.e., for a Minkowski background. In other words supersymmetry requires
\begin{equation}
\label{susy-condition-1}
D_A W=0 \qquad A=1,\ldots, b^4_{{\bf 35}^-} \,.
\end{equation}
Unlike for the four-dimensional minimal supergravity case we do not have to impose $W=0$ as well in order to obtain $V=0$. However having set the Minkowski background in the three-dimensional theory we are left with no choice and we must impose $W=0$. This is because the three-dimensional scalar curvature is proportional to the superpotential\footnote{More details can be found at the end of the second section of \cite{Becker:2003wb}.}. To summarize, the conditions that $W$ has to satisfy in order to have a supersymmetric vacua are
\begin{equation}
\label{susy-conditions}
W = 0 \quad \mathrm{and} \quad {D_A W}=0 \,,
\end{equation}
where $D_A W$ indicates the covariant derivative of $W$ with respect to the moduli fields which correspond to the metric deformations of the $Spin(7)$ holonomy manifold. Therefore if we want to break the supersymmetry of the effective three-dimensional theory, all we have to do is to impose
\begin{equation}
\label{non-susy-conditions}
W \neq 0 \quad \mathrm{or} \quad D_A W \neq 0 \,,
\end{equation}
and as we will see bellow the only way we can break the supersymmetry is through the first condition in (\ref{non-susy-conditions}) because the second condition in (\ref{susy-conditions}) is always valid. Since the general solution that emerges from our analysis invalidates the second condition in (\ref{non-susy-conditions}) from the beginning, the only way we can break the supersymmetry is to have a non-vanishing value for the vacuum expectation value of the superpotential $W$. It is obvious that once the supersymmetry is broken the relation that exists between $W$ and $D_A W$ is no longer valid and we can have for example $D_A W = 0$ and $W \ne 0$ without generating any inconsistencies. In other words the superpotential is no longer proportional to the three-dimensional scalar curvature for a non-supersymmetric theory. Having said that let us see what constraints we should impose on the internal background flux in order to obtain a supersymmetric solution.

All the fields on $M_8$ form representations of the Riemannian holonomy group $Spin(7)$. In particular, the space of differential forms on $M_8$ can be decomposed into irreducible representations of $Spin(7)$ and because the Laplace operator preserves this decomposition the de Rham cohomology groups have a similar decomposition into smaller pieces. Here we are interested in the four-form internal flux of the field strength of ${\cal M}$-theory therefore we need the decomposition of the fourth cohomology group of $M_8$
\begin{equation}
\label{general-decomp}
H^4(M_8, \mathbb{R})=H_{{\bf 1}^+}^4(M_8, \mathbb{R}) \, {\oplus} \, H^4_{{\bf 7}^+}(M_8, \mathbb{R}) \, {\oplus} \, H_{{\bf 27}^+}^4(M_8, \mathbb{R}) \, {\oplus} \, H_{{\bf 35}^-}^4(M_8, \mathbb{R}) \,.
\end{equation}
In the above expression the numerical sub-index represents the dimensionality of the representation and the $\pm$ stands for a self-dual or anti-self-dual representation. We denote by $b^n_{\bf r}$ the refined Betti number which represents the dimension of $H_{{\bf r}}^n(M_8, \mathbb{R})$. It is shown in \cite{Joyce} that for a compact manifolds which has $Spin(7)$ holonomy $b^4_{{\bf 7}^+} =0$, hence the decomposition (\ref{general-decomp}) becomes
\begin{equation}
H^4(M_8, \mathbb{R})=H_{{\bf 1}^+}^4(M_8, \mathbb{R}) \, {\oplus} \, H_{{\bf 27}^+}^4(M_8, \mathbb{R}) \, {\oplus} \, H_{{\bf 35}^-}^4(M_8, \mathbb{R}) \,.
\end{equation}
Therefore on a compact $Spin(7)$ holonomy manifold the internal flux can have three pieces
\begin{equation}
\label{F-general}
F=F_{{\bf 1}^+} \oplus F_{{\bf 27}^+} \oplus F_{{\bf 35}^-} \,.
\end{equation}
However we have showed in section \ref{eq-motion} that the most general solution which emerges from the equations of motion has to be self-dual, therefore
\begin{equation}
F_{{\bf 35}^-} = 0 \,.
\end{equation}
Although not obvious, the vanishing of the $F_{{\bf 35}^-}$ piece is related to the fact that the cosmological constant is zero in the compactified three-dimensional theory. This is because the vanishing of the $F_{{\bf 35}^-}$ piece is related to the second set of equations in (\ref{susy-conditions}). To see this let us rewrite $D_A W$ using the definition (\ref{superpotential})
\begin{equation}
D_A W = \int_{M_8} D_A \Omega \wedge F \,.
\end{equation}
The variation of the Cayley calibration with respect to the metric moduli belongs to $H_{{\bf 35}^-}^4(M_8, \mathbb{R})$ and therefore
\begin{equation}
D_A W = \int_{M_8} \xi_A \wedge F_{{\bf 35}^-} \,,
\end{equation}
hence $D_A W$ vanishes when $F_{{\bf 35}^-}$ vanishes
\begin{equation}
F_{{\bf 35}^-} = 0 \quad \Rightarrow \quad D_A W = 0 \,.
\end{equation}
In other words the general solution for the internal flux precludes a non-vanishing cosmological constant in the effective three-dimensional theory.

Regarding the first condition in (\ref{susy-conditions}) we can easily see that it is satisfied as long as $F_{{\bf 1}^+}=0$ because the Cayley calibration belongs to $H_{{\bf 1}^+}^4(M_8, \mathbb{R})$ and therefore we have that
\begin{equation}
W=\int_{M_8} \Omega \wedge F = \int_{M_8} \Omega \wedge F_{{\bf 1}^+} \,,
\end{equation}
hence
\begin{equation}
F_{{\bf 1}^+} = 0 \quad \Rightarrow \quad W = 0 \,.
\end{equation}
This means that the only piece from the internal flux that accommodates a supersymmetric vacuum is
\begin{equation}
\label{susy-flux}
F = F_{{\bf 27}^+} \,.
\end{equation}
Due to the fact that we obtain from the equations of motion that the internal flux has to be self-dual, i.e., the $F_{{\bf 35}^-}$ piece is identically zero, the only way we can break supersymmetry is by turning $F_{{\bf 1}^+}$ on in such a way that we obtain a non-vanishing value for the superpotential. It is interesting to note that breaking the supersymmetry in this way does not affect the value of the cosmological constant which remains zero.

Under these assumptions we can clearly see that the only supersymmetric solution that results from a warped compactification of ${\cal M}$-theory on a $Spin(7)$ manifold is accommodated for a Minkowski background for the external space. The same result was obtained in a previous paper \cite{Becker:2003wb} where no cosmological constant was obtained after compactification.


\section{Conclusions}
\label{conclusions}


In the present paper we have analyzed the most general warped vacua which emerge from a compactification with flux of ${\cal M}$-theory on $Spin(7)$ holonomy manifolds. The existence of the quantum corrections terms in the low energy effective action forced us to a perturbative analysis of the problem. The perturbative parameter ``$t$'' was defined in (\ref{t-def}).

We have determined that a consistent solution for the equations of motion requires that the leading order term of the internal flux has to be self-dual (\ref{self-duality}). We have also shown that the internal manifold remains Ricci flat to the $t^{-2}$ order in the perturbation theory (\ref{Ricci-t2}). By analyzing (\ref{i-t3}) we have shown that the Ricci flatness of the internal manifold is in general lost to order $t^{-3}$ and only under the restrictive condition (\ref{R-flat-condition}) the internal manifold remains Ricci flat. As a matter of fact this is the order in the perturbation theory where the influence of the quantum corrections terms is felt in the equations of motion and it is natural to expect deformations of the internal manifold to occur at this order. We have also derived a relation between the warp factor ``$A$'' and the external flux given by (\ref{warp-eflux})..

We have collected some of the properties related to the quartic polynomials in section \ref{quartic-polynomials}. In particular in section \ref{j0-computation} we have shown that $J_0$ vanishes on a $Spin(7)$ background and we have computed its first variation (\ref{Y-variation}) on a $Spin(7)$ holonomy background. We have also determined a nice formula (\ref{Y-trace}) for the trace of the first variation of $J_0$.

In section \ref{susy-solutions} we have determined the condition (\ref{susy-flux}) which has to be satisfied by the internal flux in order to obtain a supersymmetric solution. The analysis was based on the set of conditions (\ref{susy-conditions}) that were imposed to the superpotential. We have explained how a certain choice of the internal flux breaks the supersymmetry but leaves a vanishing cosmological constant.


\bigskip
\bigskip
\noindent {\Large\bf Acknowledgments}

I would like to thank Melanie Becker, Sergei Gukov and Axel Krause for useful discussions. This work was supported in part by the NSF grant PHY-0244722.


\vfill
\pagebreak


\appendix


\section{Conventions}
\label{conventions}


In what follows we present some conventions related to the Levi-Civita tensor density, some algebraic identities which involve generalized Kronecker delta and Levi-Civita symbols and a few useful $\Gamma$-matrix identities. We choose to have
\begin{equation}
\epsilon^{1 \ldots n} = \, 1 \,,
\end{equation}
and because the covariant tensor density $\epsilon_{b_1 \ldots b_n}$ is obtained from $\epsilon^{a_1 \ldots a_n}$ by lowering the indices with the help of the metric coefficients $g_{ab}$, we will have that
\begin{equation}
\epsilon_{1 \ldots n} = \, g \,,
\end{equation}
where $g=det(g_{mn})\,$. Therefore the product of two Levi-Civita symbols can be reexpressed as
\begin{equation}
\label{e-e}
\epsilon^{a_1 \ldots a_n} \, \epsilon_{b_1 \ldots b_n} = g \, \delta^{a_1 \ldots a_n}_{b_1 \ldots b_n} \,.
\end{equation}
The generalized Kronecker delta symbol which has appeared in (\ref{e-e}) is defined as
\begin{equation}
\delta^{a_1 \ldots a_n}_{b_1 \ldots b_n} = n! \delta^{a_1}_{[b_1} \ldots \delta^{a_n}_{b_n]} \,,
\end{equation}
where the antisymmetrization implies a $1/n!$ prefactor, e.g.,
\begin{equation}
\delta^a_{[m} \delta^b_{n]} = \, {1 \over {2!}} \, ( \, \delta^a_m \delta^b_n - \delta^a_n \delta^b_m \, ) \,.
\end{equation}
For a single contraction of a $(p+1)$-delta symbol in an $n$-dimensional space we have
\begin{equation}
\delta^{a_1 \ldots a_p m}_{b_1 \ldots b_p m} = (n-p) \, \delta^{a_1 \ldots a_p}_{b_1 \ldots b_p} \,,
\end{equation}
therefore in an $n$-dimensional space a $p$-delta symbol is related to an $n$-delta symbol as follows
\begin{equation}
\delta^{a_1 \ldots a_p m_1 \ldots m_{n-p}}_{b_1 \ldots b_p m_1 \ldots m_{n-p}} = (n-p)! \, \delta^{a_1 \ldots a_p}_{b_1 \ldots b_p} \,.
\end{equation}

As above, the antisymmetrization of two or more gamma matrices implies a  $1/n!$ prefactor, e.g.,
\begin{equation}
\Gamma_{mn} = \Gamma_{[m} \Gamma_{n]} = \frac{1}{2!} \, ( \, \Gamma_m\Gamma_n - \Gamma_n\Gamma_m \, ) \,.
\end{equation}
Using the fundamental relation\footnote{Written in this form the Clifford algebra and the identities (\ref{gamma-id1}) and (\ref{gamma-id2}) are  independent of the metric signature.}
\begin{equation}
\{\Gamma_m, \Gamma^n\} = 2 \delta_m^n \,,
\end{equation}
one can deduce the following gamma matrix identities
\begin{eqnarray}
&&[\Gamma_m, \Gamma^r]=2{\Gamma_m}^r \,, \nonumber \\
\label{gamma-id1}
&&\{\Gamma_{mn}, \Gamma^r \} = {2\Gamma_{mn}}^r \,, \\
&&[\Gamma_{mnp}, \Gamma^r ]=2{ \Gamma_{mnp}}^r \,, \nonumber
\end{eqnarray}
and
\begin{eqnarray}
&&\{ \Gamma_m, \Gamma^r\}=2{\delta_m}^r \,, \nonumber \\
\label{gamma-id2}
&&[\Gamma_{mn} ,\Gamma^r ]=-4{\delta^r}_{[m} \Gamma_{n]} \,, \\
&&\{ \Gamma_{mnp}, \Gamma^r\} =6 {\delta^r}_{[m} \Gamma_{np]} \nonumber \,.
\end{eqnarray}
%


\section{Useful Identities}
\label{identities}


In this appendix we derive some important formulas used in the derivations performed in section \ref{j0-computation}. At the end of the appendix we list some other useful identities providing the appropriate references for detailed explanations.

In general one has
\begin{equation}
[\nabla_m, \nabla_n]\eta=\frac{1}{4} R_{mnpq} \Gamma^{pq} \eta \,,
\end{equation}
therefore if $\eta$ is a Killing spinor then $\nabla_m \eta =0$ and we obtain the integrability condition
\begin{equation}
\label{integrability-condition}
R_{abmn} \Gamma^{mn} \eta= 0 \,.
\end{equation}
If we multiply (\ref{integrability-condition}) from the left with $\overline{\eta}\Gamma_{cd}$ we obtain
\begin{equation}
\label{id0}
R_{abmn} \overline{\eta}\Gamma_{cd}\Gamma^{mn} \eta= 0 \,.
\end{equation}
Using the identities (\ref{gamma-id1}) and (\ref{gamma-id2}), we can show that
\begin{equation}
\label{id1}
\Gamma^{a} \Gamma_{mn} = {\Gamma^{a}}_{mn} + \delta^a_m \Gamma_n - \delta^a_n \Gamma_m \,,
\end{equation}
and
\begin{equation}
\label{id2}
\Gamma^{ab} \Gamma_{mn} = {\Gamma^{ab}}_{mn} - \delta^{ab}_{mn} - 4 \delta^{[a}_{[m} {\Gamma^{b]}}_{n]} \,.
\end{equation}
If we sandwich the relation (\ref{id2}) between $\overline{\eta}$ and $\eta$ we obtain that
\begin{equation}
\label{id3}
\overline{\eta} \Gamma^{ab} \Gamma_{mn} \eta= {\Omega^{ab}}_{mn} - \delta^{ab}_{mn} \,,
\end{equation}
where
\begin{equation}
\Omega_{abmn} = \overline{\eta} \Gamma_{abmn} \eta
\end{equation}
is the Cayley calibration of the $Spin(7)$ holonomy manifold and the Killing spinor is normalized to unity, i.e., $\overline{\eta} \eta= 1$. We remind the reader that for a $Spin(7)$ holonomy manifold, terms like $\overline{\eta} \Gamma_{m_1 \ldots m_p} \eta$ are not zero only when $p=0,4$ or $8$. For details see reference \cite{Gibbons:1990er}. This is the reason why we have no contribution from the last term in (\ref{id2}). Using (\ref{id3}) we can recast (\ref{id0}) as
\begin{equation}
\label{id4}
R_{abmn} {\Omega^{mn}}_{cd} = 2 R_{abcd} \,.
\end{equation}

We have the following one index contraction between two Cayley calibrations (see for example \cite{Dundarer:1984fe} and \cite{deWit:1984gs})
\begin{equation}
\label{omega-contraction-1}
\Omega^{tabc} \Omega_{tmnp} = 6 \delta^a_{[m} \delta^b_n \delta^c_{p]} + 9 \delta^{[a}_{[m} {\Omega^{bc]}}_{np]}\,.
\end{equation}

We use the following expression for the variation of the Riemann tensor in terms of the metric fluctuations
\begin{equation}
\label{Riemann-variation}
\delta R_{abmn} = - \nabla_{[a|} \nabla_m \delta g_{|b]n} + \nabla_{[a|} \nabla_n \delta g_{|b]m} \,.
\end{equation}
The above result can be easily derived using the relation which exists between the derivative operators associated with two conformally related metrics.


\section{The Inverse Metric and Other Derivations}
\label{scaling}


In this appendix we derive the power expansion in $t$ for the inverse metric and we provide some useful relations used in the analysis of section \ref{eq-motion}. We start with the derivation of the inverse metric $g^{mn}$ followed naturally by the expansions for the Riemann tensor, the Ricci tensor and the scalar curvature that correspond to $g_{mn}$ which has $Spin(7)$ holonomy. Once we know this expansions we can perform a conformal transformation to find the corresponding tensorial quantities for the internal manifold\footnote{We have to take into account that the full metric (\ref{full-metric}) is warped.}. Also at the end of this appendix we provide the expressions for the external and internal energy-momentum tensor associated with $F_1$ and $F_2$ respectively and we list the results obtained for the term in the right hand side of (\ref{eom-g}) for the external and the internal case.

Let us consider two arbitrary square matrices $A$ and $B$ with real entries\footnote{The reader should not confuse these matrices with the warp factors.}. We want an expression for $(A + B)^{-1}$ in terms of $A$, $A^{-1}$, $B$ and $B^{-1}$. While there is no useful formula for $(A + B)^{-1}$, we can use a Neumann series to invert $A + B$ provided that $B$, for example, has small entries relative to $A$. This means that in magnitude we have
\begin{equation}
\lim_{n \, \rightarrow \, \infty} (A^{-1} \, B)^n = 0 \,.
\end{equation}
Under this assumption the inverse of $A+B$ matrix can be expressed as an infinite series
\begin{equation}
(A + B)^{-1} = \sum_{k=0}^\infty (-A^{-1}B)^k \, A^{-1} \,,
\end{equation}
which in a first approximation is given by
\begin{equation}
\label{inverse-sum}
(A + B)^{-1} = A^{-1} - A^{-1}BA^{-1} + \ldots \,.
\end{equation}
The above setup helps us to compute the inverse of the matrix $g_{mn}$ introduced in (\ref{metric-exp}). If we set $A_{mn}=t \, [g^{(1)}]_{mn}$ and $B_{mn}=[g^{(0)}]_{mn}$,  which is the ``small'' matrix\footnote{The perturbative parameter ``$t$'' was introduced in (\ref{t-def}) and in our case ``t'' is considered to be much bigger than the unity.}, then the formula (\ref{inverse-sum}) translates into
\begin{equation}
g^{mn} = t^{-1} [g^{(1)}]^{mn} + t^{-2} [g^{(2)}]^{mn} + \ldots \,,
\end{equation}
where we have defined
\begin{eqnarray}
[g^{(1)}]^{mn} &=& [g^{(1)}]^{-1}_{mn} \,, \\
{[g^{(2)}]}^{mn} &=& - [g^{(1)}]^{mp} [g^{(0)}]_{pr} [g^{(1)}]^{rn} \,,
\end{eqnarray}
and as usual $g^{mn}$ represents the inverse of $g_{mn}$.

By performing the appropriate conformal transformations we obtain the internal and external components of the eleven-dimensional Ricci tensor and the eleven-dimensional Ricci scalar that correspond to the metric (\ref{full-metric}). The results are provided in terms of the un-warped quantities, denoted here without a tilde above the symbol
\begin{eqnarray}
\label{Ricci-tensor-external}
\widetilde{R}_{\mu \nu}(M_{11}) & = & R_{\mu \nu}(M_3) - \eta_{\mu \nu} \, e^{2(A-B)} [ \, \triangle A +3 (\nabla_m A) \, (\nabla^m A) + 6 (\nabla_m A) \, (\nabla^m B) \,] \,, \\
\label{Ricci-tensor-internal}
\widetilde{R}_{mn}(M_{11}) & = & R_{mn}(M_8) - 3 \nabla_m \nabla_n A - 6 \nabla_m \nabla_n B - g_{mn} \triangle B - 3 (\nabla_m A) \, (\nabla_n A) \nonumber \\
&&+ \, 3 (\nabla_m A) \, (\nabla_n B) + 3 (\nabla_n A) \, (\nabla_m B) - 3 g_{mn} (\nabla_k A) \, (\nabla^k A) \nonumber \\
&&+ \, 6 (\nabla_m B) \, (\nabla_n B) - 6 g_{mn} (\nabla_k B) \, (\nabla^k B)\,, \\
\label{scalar-curvature}
\widetilde{R}(M_{11}) & = & e^{-2A} R(M_3) + e^{-2B} R(M_8) - e^{-2B} [ \, 6 \triangle A + 14 \triangle B \nonumber \\
&& + \, 12 (\nabla_m A) \, (\nabla^m A) + 42 (\nabla_m B) \, (\nabla^m B) + 36 (\nabla_m A) \, (\nabla^m B) \,] \,,
\end{eqnarray}
where $\triangle= \nabla^m \nabla_m$. We mention that formulas (2.34) - (2.36) in \cite{Duff:1995an} represent a generalization of the above equations\footnote{The reader must be aware of a small typo in formula (2.35) of \cite{Duff:1995an} where the term $-d \, \partial_m A \, \partial_n B$ should read $-d \, \partial_m A \, \partial_n A$.}. Because the warp factors $A$ and $B$ are of order $t^{-3}$ we will truncate (\ref{Ricci-tensor-external}), (\ref{Ricci-tensor-internal}) and (\ref{scalar-curvature}) and we will retain only the linear contributions in $A$ and $B$. The ``linearized'' expressions are
\begin{eqnarray}
\label{Ricci-tensor-external-lin}
\widetilde{R}_{\mu \nu}(M_{11})& = & R_{\mu \nu}(M_3) - \eta_{\mu \nu} \,  \triangle A + \ldots \,, \\
\label{Ricci-tensor-internal-lin}
\widetilde{R}_{mn}(M_{11}) & = & R_{mn}(M_8) - 3 \nabla_m \nabla_n A - 6 \nabla_m \nabla_n B - g_{mn} \triangle B + \ldots \,, \\
\label{scalar-curvature-lin}
\widetilde{R}(M_{11}) & = & R(M_3) + R(M_8) -  6 \triangle A - 14 \triangle B + \ldots  \,.
\end{eqnarray}

Let us compute the external and the internal components of the energy-momentum tensor associated with the field strength $F$. Because of the specific form (\ref{flux-form}) of the background flux the energy-momentum tensor defined in (\ref{en-mom-tensor}) will have the following form
\begin{eqnarray}
\label{energy-momentum-ei}
T_{\mu \nu} &=& -3 \eta_{\mu \nu} (\nabla_m f ) \, (\nabla^m f ) - {1 \over 8} \eta_{\mu \nu} F_{abmn} F^{abmn} + \ldots\,, \\
T_{mn} &=& - 6 (\nabla_m f ) \, (\nabla_n f ) + 3 g_{mn} (\nabla_p f ) \, (\nabla^p f ) \nonumber \\
&& + F_{mabp} {F_n}^{abp} - {1 \over 8} g_{mn} F_{abpr} F^{abpr} + \ldots\,,
\end{eqnarray}
where the ellipsis denote higher order terms which contain warp factors. We are also interested in computing the trace of the above tensors
\begin{eqnarray}
\eta^{\mu \nu} \, T_{\mu \nu} &=& -9 (\nabla_m f ) \, (\nabla^m f ) - {3 \over 8} F_{abmn} F^{abmn} \nonumber \\
&=& - 9 \, [g^{(1)}]^{mn} [\nabla_m f^{(2)} ] \, [\nabla_n f^{(2)} ] \, t^{-5} - {3 \over 8} [F^{(0)}]_{abmn} [F^{(0)}]^{abmn} \, t^{-4} + \ldots \,, \\
g^{mn} \, T_{mn} &=& 18 (\nabla_m f ) \, (\nabla^m f ) = 18 \, [g^{(1)}]^{mn} [\nabla_m f^{(2)} ] \, [\nabla_n f^{(2)} ] \, t^{-5} + \ldots \,,
\end{eqnarray}
where we have provided the leading order contribution of these terms. We want to note that trace of $T_{mn}$ vanishes to order $t^{-4}$ in the perturbation theory.

The external component of the left hand side of equation (\ref{eom-g}) is
\begin{equation}
- \beta {1\over \sqrt{-g}} {\delta \over \delta \eta^{\mu \nu}} \left[ \sqrt{-g} ( J_0 - {1\over 2} E_8 ) \right] = {\beta \over 4} \, \eta_{\mu \nu} \, E_8(M_8) \,,
\end{equation}
where we have used (\ref{e8-property}) and the fact that $J_0(M_{11})$ does not depend on the external metric and it vanishes on a $Spin(7)$ holonomy background\footnote{See section \ref{j0-computation} for details.}. The internal component of the left hand side of equation (\ref{eom-g}) is
\begin{equation}
- \beta {1\over \sqrt{-g}} {\delta \over \delta g^{mn}} \left[ \sqrt{-g} ( J_0 - {1\over 2} E_8 ) \right] =  - \beta \, \frac{\delta Y}{\delta g^{mn}}\,,
\end{equation}
where we have used equation (\ref{e8-variation}) and $\delta Y / \delta g^{mn}$ is given in (\ref{Y-variation}). The trace of the above equation is
\begin{equation}
\label{trace-delJ-internal}
- \beta {1\over \sqrt{-g}} \, g^{mn} \, {\delta \over \delta g^{mn}} \left[ \sqrt{-g} ( J_0 - {1\over 2} E_8 ) \right] =  - 2^{15} \, \beta \, \triangle \left[4 \, E_6(M_8)  +  \Omega \cdot z \right] \,,
\end{equation}
where we have used (\ref{Y-trace}). It is obvious that (\ref{trace-delJ-internal}) is of order $t^{-4}$ in the perturbation theory.


\vfill
\pagebreak


\bibliographystyle{hunsrt}
\bibliography{bibliography}


\end{document}